\documentclass[12pt]{article}
\usepackage{amsmath,amssymb,theorem,cite,epsfig,url,psfrag,eepic,amsmath,mathtools,mathrsfs,amsbsy,dsfont,esint,braket,cancel}

\usepackage{tikz}
\usetikzlibrary{positioning, arrows.meta}
\usepackage{comment}

\usepackage{array}

\newcolumntype{M}[1]{>{\centering\arraybackslash}m{#1}}  % центр по горизонтали и вертикали
\newcolumntype{C}[1]{>{\centering\arraybackslash}p{#1}}

\usepackage{indentfirst}
\usepackage{graphicx}
\usepackage{bm,upgreek}
\usepackage[bookmarks=true,hyperfigures=true,colorlinks=true,linkcolor=black,citecolor=black,urlcolor=black,bookmarksnumbered,hidelinks]{hyperref}
\def\XXint#1#2#3{{\setbox0=\hbox{$#1{#2#3}{\int}$}
     \vcenter{\hbox{$#2#3$}}\kern-.5\wd0}}

\advance \topmargin by -\headheight
\advance \topmargin by -\headsep     
\evensidemargin \oddsidemargin
\def\Maketitle{{\def\newpage{}\maketitle}}
\begin{document}
%\rightline{\texttt{\today}}
\title{\textbf{%On (some) differential equations (for)related to integrals of Dotsenko--Fateev type //
%Differential equations for integrals of Dotsenko--Fateev type: the case of degenerate fields $\Phi_{n,1}$ // On differential equations for integrals of Dotsenko–Fateev type in the $\Phi_{n,1}$ case
Differential equations for Dotsenko--Fateev integrals: the case of degenerate fields $\Phi_{n,1}$
}\vspace*{.3cm}}
\date{}
\author{Aleksandra Ivanova$^{1}$ 
\\[\medskipamount]
\parbox[t]{0.85\textwidth}{\normalsize\it\centerline{1. HSE University, 6 Usacheva str., Moscow 119048, Russia}}
%\\
%\parbox[t]{0.85\textwidth}{\normalsize\it\centerline{2. Krichever Center, Skolkovo Institute of Science and Technology, 121205 Moscow, Russia}}
%\\
%\parbox[t]{0.85\textwidth}{\normalsize\it\centerline{3. Landau Institute for Theoretical Physics, 142432 Chernogolovka, Russia}}
}
\Maketitle
\begin{abstract}
We study higher-order differential equations  corresponding to integrals of the Dotsenko--Fateev type and associated with four-point correlation functions with degenerate fields $\Phi_{n,1}$. We determine the general structure and coefficients of the equations in terms of the integral parameters $ \{ A,B,C,g \} $.
%We derive recurrence relations that determine the coefficients of these differential equations of arbitrary order in terms of the integral parameters 
In particular, special cases of these equations in CFT are used for verification. In addition, the relationship between higher-order differential equations and hypergeometric operators is established.
\end{abstract}
%%%%%%%%%%%%%%%%%%%%%%%%%%%%%%%%%%%%%%%%%%%%%%%%%%%%%%%%%%%%%%%%%%%%%%%%%%%%%%%%%%%%%%%%%%%%%%%%%
%\tableofcontents
\section{Introduction}
One of the main problems in Conformal Field Theory (CFT) is the calculation of correlation functions. In particular, a four-point correlation function that includes at least one degenerate field $\Phi_{n,m}$ can be obtained as a solution of a differential equation of order $nm$, known as the BPZ equation \cite{belavin1984infinite}. Studying Coulomb gas systems, Dotsenko and Fateev obtained the integral representation of the solutions for four-point correlation functions \cite{dotsenko1984conformal, dotsenko1985four}. This was obtained using a method based on screening operators and charge balance conditions. This approach is more universal and simpler than directly solving the corresponding differential equations.

Similar ideas were expressed in the context of Liouville field theory (LFT) \cite{Polyakov:1981rd}, which is defined by the Lagrangian
\begin{align}
    \mathcal{L} = \frac{1}{4\pi} (\partial \varphi)^2 + \mu e^{2 b \varphi},
\end{align}
where $b$ is the coupling constant related to the central charge $c_L$ and the background charge $Q$ in the following way:
\begin{align*}
    c_L = 1+ 6 Q^2  , \hspace{20pt} Q = b + \frac{1}{b}.
\end{align*}
It was shown that the interacting theory for $\mu \neq 0$ can be studied using free field methods \cite{goulian1991correlation,Zamolodchikov:1995aa}. Moreover, multipoint correlation functions with poles admit Coulomb gas integral representations for the corresponding residues. In \cite{fateev2006coulomb}, it was shown that the holomorphic part of the four-point correlation function with one degenerate field $V_{n,1}$ in LFT, up to a factor, takes the following form:
\begin{align}
\begin{split} \label{nn2}
    &\left\langle V_{-\frac{(n-1) b}{2}}(z)V_{\alpha_1}(0)V_{\alpha_2}(1)V_{\alpha_3}(\infty)\right\rangle_{\text{hol.}}
    \sim z^{(n-1)b\alpha_1} (z - 1)^{(n-1)b\alpha_2} \cdot \\
    &\hspace{9cm} \cdot \int \prod_{k=1}^{n-1} dt_k t_k^{A} (t_k - 1)^{B} (t_k - z)^{C} \prod_{i<j} (t_i-t_j)^{g} ,
\end{split}
\end{align}
where
\begin{align} \label{n2}
\begin{split}
    A &= b \Big(- \alpha_1 + \alpha_2 + \alpha_3 - Q + \frac{(n-1) b}{2} \Big), \hspace{10pt}
    B = b \Big( \alpha_1 - \alpha_2 + \alpha_3 - Q + \frac{(n-1) b}{2} \Big), \\
    C &= b \Big( Q + \frac{(n-1) b}{2} - (\alpha_1+\alpha_2+\alpha_3) \Big), \hspace{17pt} g = -2b^2.
\end{split}
\end{align}
As in the formula above, in what follows we consider only the holomorphic part of correlation functions and differential equations. The antiholomorphic part is obtained by complex conjugation, as a consequence of the factorization of two-dimensional conformal symmetry. This type of integral is often defined using multiple Pochhammer contours. This choice of contour allows one to analytically continue the integral to a wider domain of the parameters $\{A, B, C, g\}$, thereby removing the convergence restrictions associated with real contour integration. In the holomorphic setting considered here, this procedure is directly applicable.

To verify that the integral representation indeed satisfies the BPZ equations, we start with the simplest nontrivial cases. The first non-trivial example is the second-order equation, associated with the degenerate fields $\Phi_{2,1}$ or $\Phi_{1,2}$\footnote{$\Phi_{n,m}$ is the rename of the primary field $V_{n,m}$ in Liouville CFT.}. This equation reduces to the Gauss hypergeometric equation
\begin{align}
    z(z-1) F_2''(z) - \big( (A + C)(z-1) + (B + C) z \big) F_2'(z) + C (A + B + C + 1) F_2(z) = 0,
\end{align}
and its solution admits an integral representation via Euler's formula
\begin{align} \label{n5}
    F_2(z) = \int\limits_{C_1} dt_1 t_1^A (t_1 - 1)^B (t_1 - z)^C.
\end{align}
For $n = 2$, the general integral form of the correlation function (\ref{nn2}) is reduced to (\ref{n5}), up to a factor.

It would be natural to assume that higher-order BPZ equations are also reducible to hypergeometric form, but this is not the case. The third-order equation, corresponding to $\Phi_{3,1}$ or $\Phi_{1,3}$, was explicitly derived by Dotsenko and Fateev \cite{dotsenko1984conformal}:
\begin{align} \label{1}
\begin{split}
    &z^2 (z-1)^2 F_3'''(z) + \big( K_1 z + K_2 (z-1) \big) z (z-1) F_3''(z) + \\
    &+\big( L_1 z^2 + L_2 (z - 1)^2 + L_3 z(z-1) \big) F_3'(z) + (M_1 z + M_2 (z-1)) F_3(z) = 0,
\end{split}
\end{align}
where
\begin{align*}
    K_1 =& - (g + 3 B + 3 C), \hspace{70pt}
    K_2 = -(g + 3 A + 3 C), \\
    L_1 =& (B + C) (2 B + 2 C + g + 1), \hspace{23pt}
    L_2 = (A + C) (2 A + 2 C + g + 1), \\
    L_3 =& (B + C) (2 A + 2 C + g + 1) + (A + C) (2 B + 2 C + g + 1) + \\
    &+(C - 1) (A + B + C) + (3 C + g) (A + B + C + g + 1), \\
    M_1 =& -C (2 B + 2 C + g + 1) (2 A + 2 B + 2 C + g + 2), \\
    M_2 =& -C (2 A + 2 C + g + 1) (2 A + 2 B + 2 C + g + 2).
\end{align*}
The corresponding solution is expressed as a double integral 
\begin{align} \label{2}
    F_3(z) = \int\limits_{C_1} dt_1 \int\limits_{C_2} dt_2 t_1^A (t_1 - 1)^B (t_1 - z)^C  t_2^A (t_2 - 1)^B (t_2 - z)^C (t_1 - t_2)^g,
\end{align}
where the set of parameters $\{ A,B,C,g \}$ can be associated with (\ref{n2}). This is an example of a Fuchsian differential equation with three singular points at $0$, $1$, and $\infty$. Moreover, the Dotsenko--Fateev equation is a generalization of the hypergeometric case. Thus, when $2B+2C+g+1=0$, both $L_1$ and $M_1$ vanish, and the resulting equation corresponds to the hypergeometric
operator associated with \\ ${}_3 F_2 (-2C,-1-2A, B-A; -A-C, 2B-2A ;z)$ \cite{mimachi2024irreducibility}.

An important aspect of the study of differential equations is the analysis of their solutions. This naturally leads to the monodromy problem for such objects. This issue was considered from the point of view of the physical requirement that the correlation functions be single-valued. The question of convergence of such integrals has been addressed in detail in \cite{dotsenko1984conformal,mimachi2024irreducibility}. These considerations can be easily generalized to the case of $n$-dimensional integral solutions \cite{dotsenko1985four, mimachi2007connection}.

The question of a generalized version of the Dotsenko--Fateev equation may arise. Some attempts have been made using the Riemann--Liouville transformation and middle convolution \cite{ebisu2023study,haraoka2021shift,haraoka2013prolongability}. However, such results are unusual in the context of CFT. In this theory, it is desirable to find equations that are satisfied by the Dotsenko--Fateev integrals \cite{dotsenko1985four}. In \cite{vishnevskaya2015differential}, a fourth-order differential equation is presented whose form is postulated on the basis of an ansatz. The verification is performed using the method of undetermined coefficients for the following integral solution:
\begin{align}
    F_4(z) =  \int \limits_{C_1} dt_1 \int \limits_{C_{2}} dt_{2} \int \limits_{C_{3}} dt_{3} \prod \limits_{i=1}^{3} t_i^A (t_i - 1)^B (t_i - z)^C \prod \limits_{i<j} (t_i - t_j)^g.
\end{align}
However, this approach does not reveal the general structure and does not suggest systematic patterns. Although Dotsenko and Fateev did not explicitly demonstrate the correspondence between (\ref{1}) and (\ref{2}), this was later established by Mimachi in \cite{mimachi2024irreducibility}. The method relies on combinations of total derivatives. In our work, we extended this approach to obtain higher-order equations for similar solutions. In contrast to earlier work, our construction allows an arbitrary number of integration variables $t_{i}$ and does not impose any restrictions between the parameters $\{ A,B,C,g \}$. 

The general differential equations obtained in this work allow us to verify the correspondence between the Dotsenko--Fateev integrals and the BPZ equations. This correspondence has been studied by Zhu within the probabilistic framework of Liouville theory using Gaussian multiplicative chaos \cite{zhu2020higher}. In the present work, we adopt a different approach similar to \cite{mimachi2024irreducibility}, working directly within the standard CFT formalism based on the construction of total derivatives.

The paper is organized as follows. In section 2, we give a brief overview of the method used in \cite{mimachi2024irreducibility}. In section 3, we consider the generalization of this algorithm. Explicit expressions for the differential equation are given. In addition, a method for obtaining the coefficients of these equations is presented. After this, a verification of such equations is given with special cases arising in CFT. In section 4, we discuss the relations between general differential equations and hypergeometric cases.

\section{Verification of the third-order differential equation}\label{sec: Chapter 2}
To verify that a function satisfies a differential equation, one typically substitutes the function into the equation. For an integral representation of a solution, this substitution yields a sum of total derivatives. The verification therefore reduces to examining the boundary contributions: if every term vanishes at the boundary, the integral representation indeed defines a solution. In \cite{mimachi2024irreducibility}, the reverse procedure—deriving equation~(\ref{1}) from the solution~(\ref{2})—is presented as follows.

Let us define $\mathcal{I}_3(z)$ as the integrand of (\ref{2}) and introduce the following functions:
\begin{align} \label{3}
\begin{split}
    &p_1 = \frac{1}{t_1 - z} + \frac{1}{t_2 - z}, \hspace{2.65cm}
    e_2 = \frac{1}{(t_1 - z)(t_2 - z)}, \\
    &p_2 = \left( \frac{1}{t_1 - z} \right)^2 + \left( \frac{1}{t_2 - z} \right)^2, \hspace{1cm}    e_{(2,1)} = \frac{1}{(t_1 - z)(t_2 - z)} \left( \frac{1}{t_1 - z} + \frac{1}{t_2 - z} \right).
\end{split}
\end{align}
Their combinations with the integrand allow us to relate them to the derivative of $\mathcal{I}_3(z)$
\begin{align} \label{4}
\begin{split}
    &p_1 \mathcal{I}_{3}(z) = -\frac{1}{C} \partial_z \mathcal{I}_{3}(z), \\
    &(1 - C) p_2 \mathcal{I}_{3}(z) = -\frac{1}{C} \partial_
    z^2 \mathcal{I}_{3}(z) + 2 C e_2 \mathcal{I}_{3}(z), \\
    &(1 - C) e_{(2,1)} \mathcal{I}_{3}(z) = \partial_z (e_2 \mathcal{I}_{3}(z)).
\end{split}
\end{align}
However, one of them cannot be expressed in terms of derivatives of the integrand and remains an unknown function. Thus, we need an additional condition to find $e_2 \mathcal{I}_{3}(z)$. 

To solve this issue, it is useful to consider some combination of total derivatives associated with the integral solution. It is easy to show that the following decomposition is valid:
\begin{align} \label{5}
\begin{split}
   & \partial_{t_1} \left[ \frac{t_1(t_1 - 1)}{t_1 - z} \mathcal{I}_{3}(z) \right] = \\
   &\Bigg( (A + B + C + 1) + \frac{(z-1)(A + C) + z(B + C)}{t_1 - z} + \frac{z(z-1)(C - 1)}{(t_1 - z)^2} + \frac{g t_1(t_1 - 1)}{(t_1 - z)(t_1 - t_2)} \Bigg) \mathcal{I}_{3}(z).
\end{split}
\end{align}
The expression for the variable $t_2$ is similar. One can see that the terms in (\ref{5}) correspond to (\ref{3}), except for the last one. Fortunately, this term can also be reduced to the form of the above functions using the following identity
\begin{align*}
    \frac{t_1(t_1 - 1)}{(t_1 - z)(t_1 - t_2)} - \frac{t_2(t_2 - 1)}{(t_2 - z)(t_1 - t_2)} = 1 + z(1 - z) e_2.
\end{align*}
It is the reason why we must consider the sum of similar type total derivatives. As a result, we obtain the following equality:
\begin{align} \label{6}
\begin{split}
    &\partial_{t_1} \left[ \frac{t_1(t_1 - 1)}{t_1 - z} \mathcal{I}_{3}(z) \right] + \partial_{t_2} \left[ \frac{t_2(t_2 - 1)}{t_2 - z} \mathcal{I}_{3}(z) \right] = \Bigg( (2A + 2B + 2C + g + 2)+ \\
    &\hspace*{90pt}+ \big((z-1)(A + C) + z(B + C) \big) p_1 + z(z-1)(C - 1) p_2 - z(z-1) g e_2 \Bigg) \mathcal{I}_{3}(z) .
\end{split}
\end{align}
Consequently,
\begin{align} \label{7}
\begin{split}
    &e_2 \mathcal{I}_{3}(z) = \frac{1}{z(1 - z)(2C + g)} \Bigg( \partial_{t_1} \left[ \frac{t_1(t_1 - 1)}{t_1 - z} \mathcal{I}_{3}(z) \right] +  \partial_{t_2} \left[ \frac{t_2(t_2 - 1)}{t_2 - z} \mathcal{I}_{3}(z) \right] -  \\
    &\hspace{16pt} -(2A + 2B + 2C + g + 2) \mathcal{I}_{3}(z) + \frac{(z-1)(A + C) + z(B + C)}{C} \partial_z \mathcal{I}_{3}(z)  - z(z-1) \frac{1}{C} \partial_z^2 \mathcal{I}_{3}(z) \Bigg) .
\end{split}
\end{align}
Thus, up to total derivatives, all functions in (\ref{3}) are expressed in terms of $\mathcal{I}_{3}(z)$ and its derivatives. 

We can go further and consider the equality obtained from the following total derivatives:
\begin{align} \label{8}
\begin{split}
   & \partial_{t_i} \left[ \frac{t_i(t_i - 1)}{(t_i - z)(t_j - z)} \mathcal{I}_{3}(z) \right] = \\
   &\Bigg( \frac{(A + B + C + 1)}{(t_j - z)} + \frac{(z-1)(A + C) + z(B + C)}{(t_i - z)(t_j - z)} + \frac{z(z-1)(C - 1)}{(t_i - z)^2 (t_j - z)} + \frac{g t_i(t_i - 1)}{(t_i - z)(t_j - z)(t_i - t_j)} \Bigg) \mathcal{I}_{3}(z).
\end{split}
\end{align}
Taking into account
\begin{align*}
      \frac{1}{(t_1 - z)(t_2 - z)} \Bigg( \frac{t_1(t_1 - 1)}{(t_1 - t_2)} - \frac{t_2(t_2 - 1)}{(t_1 - t_2)} \Bigg) = p_1 + (2z - 1)e_{2},
\end{align*}
the key equation may be obtained
\begin{align} \label{9}
\begin{split}
    \partial_{t_1} &\left[ \frac{t_1(t_1 - 1)}{(t_1 - z)(t_2 - z)} \mathcal{I}_{3}(z) \right] + \partial_{t_2} \left[ \frac{t_2(t_2 - 1)}{(t_1 - z)(t_2 - z)} \mathcal{I}_{3}(z) \right] = \\
    &\hspace{10pt}= - \frac{ (A + B + C + g + 1)}{C} \partial_z \mathcal{I}_{3}(z) + \Big( 2 \big( (z-1)(A + C) + z(B + C) \big) + (2z - 1)g \Big) e_2 \mathcal{I}_{3}(z) - \\
    &\hspace{23pt} - z(z-1) \partial_z ( e_2 \mathcal{I}_{3}(z) ).
\end{split}
\end{align}
It is easy to see that the new independent equation above contains the auxiliary term $e_2 \mathcal{I}_{3}(z)$ and its derivative, which cannot be eliminated using (\ref{4}). Therefore, considering (\ref{6}) is essential. 

The substitutions of (\ref{4}) and (\ref{7}) in (\ref{9}), together with integration, lead to the Dotsenko--Fateev equation (\ref{1}). It is important to recall that, in our framework, integrals of total derivatives vanish due to boundary conditions in our consideration.

%\section{Generalized form of equations and special cases in conformal field theory}
\section{Generalization to higher-order differential equations}
In \hyperref[sec: Chapter 2]{the previous section}, it was shown that the introduced functions (\ref{3}) and certain combinations of total derivatives ((\ref{6}) and (\ref{9})) lead to the system of equations that allows us to obtain the differential equation corresponding to the solution $F_3(z)$. In this section, we generalize this method for constructing differential equations for solutions of the following type:
\begin{align}
    F_n(z) =  \int\limits_{C_1} dt_1 \dots \int\limits_{C_{n-1}} dt_{n-1} \prod \limits_{i=1}^{n-1} t_i^A (t_i - 1)^B (t_i - z)^C \prod \limits_{i<j} (t_i - t_j)^g.
\end{align}

For simplicity, we first prepare a preliminary list of auxiliary functions. After that we move on to a detailed explanation and consideration of the generalized system of equations associated with total derivatives. 

In general, auxiliary functions that are symmetric in variables $t_i$ have the following forms:
\begin{align} \label{11}
\begin{split}
    &m_i(t_1,\dots,t_{n-1}) =  \frac{1}{(t_1 - z)^i} + \dots, \hspace{1 cm} m_{ij}(t_1,\dots,t_{n-1}) = \frac{1}{(t_1 - z)^i (t_2 - z)^j} + \dots, \\
     &m_{ijk}(t_1,\dots,t_{n-1}) = \frac{1}{(t_1 - z)^i (t_2 - z)^j (t_3 - z)^k} + \dots, \\
     &\dots\\
      &m_{ijk \dots l}(t_1,\dots,t_{n-1}) = \frac{1}{(t_1 - z)^i (t_2 - z)^j (t_3 - z)^k \cdot \dots \cdot (t_{n-1} - z)^l} + \dots 
\end{split}
\end{align}
We can notice that the functions (\ref{11}) are connected with each other via $\partial_z$
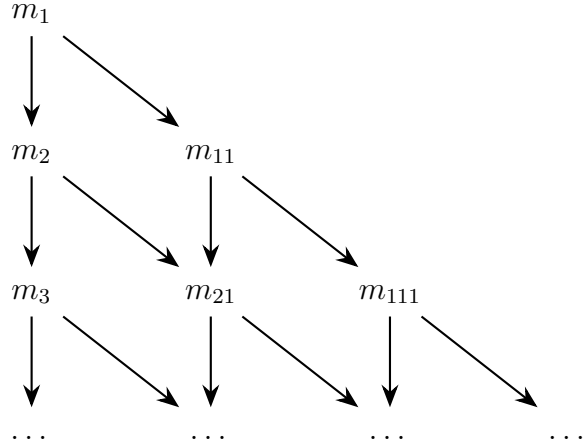
\begin{center}
\begin{tikzpicture}[
    node distance=1.2cm,
    arr/.style={-{Stealth[length=3mm]}, thick, black},
    every node/.style={text height=1.5ex, text depth=0.5ex, minimum width=2.8em} 
    ]

% 1 ряд 
\node (p1) {$m_1$};

% 2 ряд 
\node (p2) [below=of p1] {$m_2$};
\node (e11) [right=of p2] {$m_{11}$};

% 3 ряд
\node (p3) [below=of p2] {$m_3$};
\node (e21) [below=of e11] {$m_{21}$};
\node (e111) [right=of e21] {$m_{111}$};

% 4 ряд
\node (p4) [below=of p3] {$\dots$};
\node (e31) [below=of e21] {$\dots$};
\node (e211) [below=of e111] {$\dots$};
\node (e1111) [right=of e211] {$\dots$};

% Стрелка 1 ряд
\draw[arr] (p1) -- (p2);
\draw[arr] (p1) -- (e11);

% Стрелка 2 ряд
\draw[arr] (p2) -- (p3);
\draw[arr] (p2) -- (e21);
\draw[arr] (e11) -- (e21);
\draw[arr] (e11) -- (e111);

% Стрелка 3 ряд
\draw[arr] (p3) -- (p4);
\draw[arr] (p3) -- (e31);
\draw[arr] (e21) -- (e31);
\draw[arr] (e21) -- (e211);
\draw[arr] (e111) -- (e211);
\draw[arr] (e111) -- (e1111);

\end{tikzpicture}
\end{center}
For reasons that will be mentioned later, we will consider part of (\ref{11}) that is related the following way:
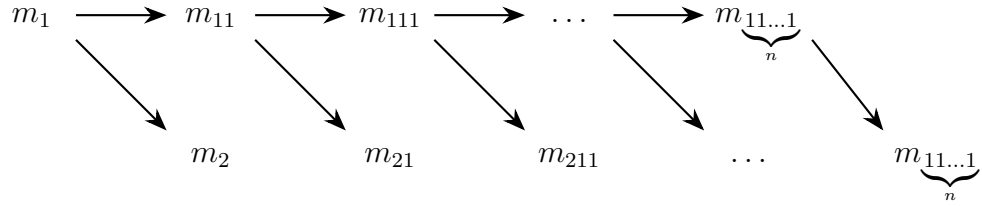
\begin{center}
\begin{tikzpicture}[
    node distance=1.2cm,
    arr/.style={-{Stealth[length=3mm]}, thick, black},
    every node/.style={text height=1.5ex, text depth=0.5ex, minimum width=2.8em} 
]

% 1 ряд 
\node (p1) {$m_1$};
\node (e11) [right=of p1] {$m_{11}$};
\node (e111) [right=of e11] {$m_{111}$};
\node (e1111) [right=of e111] {$\mathstrut \dots$};
\node (e1111n) [right=of e1111] {$m_{\smash{\underbrace{\scriptstyle 11\ldots 1}_{n-1}}} $};
%\node (ent) [right=of e1111n] {$e_{\smash{\underbrace{\scriptstyle 11\ldots 1}_{n}}} $};

% 2 ряд 
\node (p2) [below=of e11] {$m_2$};
\node (e21) [below=of e111] {$m_{21}$};
\node (e211) [below=of e1111] {$m_{211}$};
\node (e211n) [right=of e211] {$\mathstrut \dots$};
\node (e211nn) [right=of e211n] {$m_{2{\smash{\underbrace{\scriptstyle 11\ldots 1}_{n-2}}}} $};
%\node (e211ent) [right=of e211nn] {$e_{\smash{\underbrace{\scriptstyle 21\ldots 1}_{n}}}$};

% 3 ряд
%\node (p3) [below=of e21] {$\dots$};
%\node (p3n) [below=of e211] {$\dots$};
%\node (p3nn) [below=of e211n] {$\dots$};

% Стрелка вправо: 1 ряд
\draw[arr] (p1) -- (e11);
\draw[arr] (e11) -- (e111);
\draw[arr] (e111) -- (e1111);
\draw[arr] (e1111) -- (e1111n);
%\draw[arr] (e1111n) -- (ent);

% Стрелка вправо: 2 ряд
%\draw[arr] (p2) -- (e21);
%\draw[arr] (e21) -- (e211);
%\draw[arr] (e211) -- (e211n);
%\draw[arr] (e211n) -- (e211nn);
%\draw[arr] (e211nn) -- (e211ent);

% Стрелка вправо: 3 ряд

% Стрелка вниз по диагонали: из 1 во 2 ряд
\draw[arr] (p1.south east) -- (p2.north west);
\draw[arr] (e11.south east) -- (e21.north west);
\draw[arr] (e111.south east) -- (e211.north west);
\draw[arr] (e1111.south east) -- (e211n.north west);
\draw[arr] (e1111n.south east) -- (e211nn.north west);
%\draw[arr] (ent.south east) -- (e211ent.north west);

% Стрелка вниз по диагонали: из 2 в 3 ряд
%\draw[arr] (p2.south east) -- (p3.north west);
%\draw[arr] (e21.south east) -- (p3n.north west);
%\draw[arr] (e211.south east) -- (p3nn.north west);

\end{tikzpicture}
\end{center}
The scheme above helps us to notice that some functions in this set can be expressed similarly to (\ref{4})
\begin{align}
\begin{split}
    m_1 \mathcal{I}_n(z) &= -\frac{1}{C} \partial_z \mathcal{I}_n(z), \\
    m_2 \mathcal{I}_n(z) &= - \frac{1}{(C - 1)} \partial_z (m_1 \mathcal{I}_n(z)) - 2 \frac{C}{(C - 1)} m_{11} \mathcal{I}_n(z), \\
    m_{21} \mathcal{I}_n(z) &= -\frac{1}{(C - 1)} \partial_z (m_{11} \mathcal{I}_n(z)) - 3 \frac{C}{(C - 1)} m_{111} \mathcal{I}_n(z), \\
    %e_{211} F_n(z) &= -\frac{1}{(C - 1)} \partial_z (e_{111} F_n(z)) - 4 \frac{C}{(C - 1)} e_{1111} \mathcal{I}_n(z) \\
    &\dots \\
    m_{\smash{\underbrace{\scriptstyle 21\ldots 1}_{n-2}}} \mathcal{I}_n(z) &= -\frac{1}{(C - 1)} \partial_z (m_{\smash{\underbrace{\scriptstyle 11\ldots 1}_{n-2}}} \mathcal{I}_n(z)) - (n-1) \frac{C}{(C - 1)} m_{\smash{\underbrace{\scriptstyle 11\ldots 1}_{n-1}}} \mathcal{I}_n(z), \\
    m_{\smash{\underbrace{\scriptstyle 21\ldots 1}_{n-1}}} \mathcal{I}_n(z) &= -\frac{1}{(C - 1)} \partial_z (m_{\smash{\underbrace{\scriptstyle 11\ldots 1}_{n-1}}} \mathcal{I}_n(z)),
\end{split}
\end{align}
where $\mathcal{I}_n(z)$ is the integrand of $F_n(z)$. This procedure makes it possible to reduce the number of auxiliary functions that are required to construct a differential equation by a factor of two. The remaining goal is to find combinations of the form $m_{1 \dots 1} \mathcal{I}_n(z)$. They can be obtained from the system of equations that will be discussed in the following, and these combinations will be presented  in terms of partial derivatives of $\mathcal{I}_n(z)$.

Now, let us turn to the system of equations. There are many equations that can be obtained from combinations of total derivatives such as (\ref{6}) and (\ref{9}), but with other degrees in $(t_1 - z)^i \cdot ...\cdot (t_{n-1} - z)^j$ in the denominator. But it turns out that some part of them is dependent. The expansions of some total derivatives can be expressed in terms of others. This is due to the fact that the functions in (\ref{11}) are connected through $\partial_z$. There are different ways to select an independent set, but we provide one that is simpler and easier to generalize. For this reason, it is convenient to consider the decomposition of the total derivatives expressed in terms of $\{m_{2 \dots 1}, m_{1 \dots 1} \}$. All these arguments can be easily followed from the schemes above. The desired independent system is formed from the following elements:
\begin{align*}
    I_1 &= \sum_{i=1}^{n-1} \partial_{t_i} \Bigg[ \frac{t_i (t_i - 1)}{(t_i - z)} \mathcal{I}_n(z) \Bigg] , \\
    I_2 &= \sum_{i=1}^{n-1} \partial_{t_i} \Bigg[ \frac{t_i (t_i - 1)}{(t_i - z)} m_1(\dots, \check{t_i}, \dots) \mathcal{I}_n(z) \Bigg] ,\\
    &\dots \\
    I_{n-1} &= \sum_{i=1}^{n-1} \partial_{t_i} \Bigg[ \frac{t_i (t_i - 1)}{(t_i - z)} m_{\smash{\underbrace{\scriptstyle 1\ldots 1}_{n-2}}}(\dots, \check{t_i}, \dots) \mathcal{I}_n(z) \Bigg],
\end{align*}
where $\check{t_i}$ means that this variable is absent as an argument in functions. The expansion $I_i$ in terms of the additional functions can be written in general form
\begin{align} \label{13}
    I_i =& R(i) f_{i-1} + (W_1(i)(z - 1) + W_0(i)z) f_{i} - z(z-1)f'_{i} - z(z-1) S(i) f_{i+1}, \hspace{15pt} 1 \leq i \leq n-1,
\end{align}
where we introduce factors
\begin{align*}
    &R(i) = (n - i)(A + B + C + 1) + ( T_{n - 2} - T_{i - 2} ) g, \\
    &W_1(i) = i (A + C) + T_{i - 1} g, \hspace{10 pt}
    W_0(i) = i (B + C) + T_{i - 1} g, \\
    &S(i) = (i + 1) C + T_{i } g,
\end{align*}
and define the function
\begin{align*}
    f_l = 
    \begin{cases}
    \mathcal{I}_n(z) , \hspace{46pt} l = 0 \\
    m_{\smash{\underbrace{\scriptstyle 1\ldots 1}_{l}}} \mathcal{I}_n(z),  \hspace{14pt} 0 < l < n\\
    0,  \hspace{68pt} l = n \\
    \end{cases}
\end{align*}
In the coefficients above, we used the triangular number $T_{n} = C^2_{n+1}$, which is essentially a redefinition of the binomial coefficient and allows us to avoid ambiguity in notation. It is easy to notice that these equations (\ref{13}) can be considered as recurrence relations for the $f$-functions
\begin{align}
    f_{i+1} = \frac{1}{z(z-1) S(i)} R(i) f_{i-1} + \frac{1}{z(z-1) S(i)} ( W_1(i) (z-1) + W_0(i) z ) f_{i} - \frac{1}{S(i)} f'_{i} + \dots
\end{align}
where the last term corresponds to total derivatives and does not contribute after integration. Information about the new element $f_{i+1}$ is contained at each step $i$ except the last one, $i = n-1$, since $f_n = 0$. All this allows us to express unknown combinations $m_{1 \dots 1} \mathcal{I}_n$ in terms of partial derivatives of the integrand. Finally, substitution of these expressions into the last equation $I_{n-1}$ and integration allow us to find the desired differential operator for $F_n(z)$. At the same time, it is important to mention that the recursion is constructed correctly because the initial functions $f_0$ and $f_1$ are expressed in terms of $\mathcal{I}_n(z)$ and $\partial_z \mathcal{I}_n(z)$ at the beginning. 

As a result, we obtain the following differential equation:
\begin{align} \label{15}
    \sum_{k = -1}^{n - 1} \Big( \mathcal{W}_{n-k}(0) + z(z-1) \mathcal{R}_{n-k}(0) \Big) (-1)^k z^k (z-1)^k F_n^{(k+1)}(z) =0,
\end{align}
where the coefficients are defined by the recurrence relations
\begin{align*}
    &\mathcal{W}_{1}(i) = 1, \hspace{20pt} \mathcal{W}_{2}(i) = \sum_{k = i+1}^{n-1} \Big( (z-1) W_1(k) + z W_0(k) \Big), \\
    &\mathcal{W}_{l}(i) = \sum_{k= 1}^{l-1} \big((n-1)-(i+l-2) \big)^{\overline{k-1}} \big( W_1(i+1)(z-1)^k + W_0(i+1)z^k \big) \mathcal{W}_{l-k}(i + 1) + \\
    &\hspace{9cm} + \big( z(z-1) \mathcal{R}_l(i+1) + \mathcal{W}_l(i+1) \big), \hspace{20pt}l \geq 3\\
    &\mathcal{W}_{n+1}(0) = \mathcal{R}_{1}(0) = \mathcal{R}_{2}(0) = 0,\\
    &\mathcal{R}_{3}(i) = S(i) R(i+1) \mathcal{W}_{1}(i), \\
    &\mathcal{R}_{4}(i) = S(i) R(i+1) \Big( \mathcal{W}_{2}(i+1) + \big( (n-1) - (i+1) \big) \big( z + (z-1) \big) \Big),\\
    &\mathcal{R}_{l}(i) = S(i) R(i+1) \sum_{k = 2}^{l-1} \big((n-1)-(i+l-3) \big)^{\overline{k-2}} \big( z^{k-1} - (z-1)^{k-1} \big) \mathcal{W}_{l-k}(i+1),  \hspace{20pt}l \geq 5,
\end{align*}
where $x^{\overline{k}}  = x(x+1)\dots (x+(k-1))$ is the Pochhammer symbol. To correctly define recurrence relations, we must specify their initial values. It is necessary and sufficient to define the starting functions $\mathcal{W}$ as $\mathcal{W}_l(n-l+1) =0$.

At the same time, equations (\ref{15}) can be reduced to canonical form. In particular, their coefficients can be represented as a sum of terms of the form $(z-1)^i z^j$
\begin{align} \label{16}
     \sum_{q = -1}^{n - 1} \underset{l+m+1=n-q}{\underset{l, m \geq 0}{\sum}}\Big(\mathcal{Z}_{n-q}(l,m,0) (z-1)^l z^m \Big) (-1)^q z^q (z-1)^q F_n^{(q+1)}(z) =0, 
\end{align}
where the coefficients $\mathcal{Z}_{k}(l,m,i)$ with the restriction $k=l+m+1$ have the following form
\begin{align*}
    \mathcal{Z}_{1}(l,m,i) &= 1,\\
    \mathcal{Z}_{2}(l,m,i) &= \sum_{k=i+1}^{n - 1} ( (1 - \delta_{l,0}) W_1(k) + (1 - \delta_{m,0}) W_0(k) ),\\
    \mathcal{Z}_{k}(l,m,i) &= (1 - \delta_{k, n + 1} ) Z^W_k(l, m, i) + \\ 
    &\hspace{20pt} + ( 1 - \delta_{l,0} ) ( 1 - \delta_{m,0} ) S(i) R(i+1) Z^R_{k-2}(l-1, m-1, i+1) , \hspace{20pt} 3 \leq k \leq n+1
\end{align*}
are defined by the functions
\begin{align*}
    &Z^W_1(l, m, i) =  Z^R_1(l, m, i) = 1 ,\\
    &Z^W_2(l, m, i) = \sum_{k=i+1}^{n - 1} \Big( (1 - \delta_{l,0}) W_1(k) + (1 - \delta_{m,0}) W_0(k) \Big), \\
    &Z^W_k(l, m, i) = W_1(i+1) \sum_{K=0}^{l-1} \big( (n-1) - (i+k-2) \big)^{\overline{l-1-K}} Z^W_{1+m+K}(K, m, i+1) +\\
    &\hspace{60pt} + W_0(i+1) \sum_{K=0}^{m-1} \big( (n-1) - (i+k-2) \big)^{\overline{m-1-K}} Z^W_{1+l+K}(l, K, i+1) + \\
    &\hspace{60pt} + Z^W_k(l, m, i+1) + \\
    &\hspace{60pt} + (1 - \delta_{l,0})(1 - \delta_{m,0}) S(i+1) R(i+2) Z^R_{k-2}(l-1, m-1, i+2), \hspace{20pt} k \geq 3,\\
    &Z^R_2(l, m, i) = \sum_{k=i+1}^{n - 1} \Big( (1 - \delta_{l,0}) \bigl( W_1(k) + 1 \bigr) + (1 - \delta_{m,0}) \bigl( W_0(k) + 1 \bigr) \Big)\\
    &Z^R_k(l, m, i) = \sum_{q=1}^{k} \big( (n -1) - (i+k-2) \big)^{\overline{k-q}} \sum_{L = \max(0, q-m-1)}^{\min(l, q-1)} Z^W_q(L, q-L-1, i), \hspace{20pt} k \geq 3
\end{align*}
It is important to emphasize that for the correct determination of the factors, the base of the recursion must be defined as $Z^W_k(l,m,n+1-k)=0$. 

Let us comment on the structure of the coefficients in the equation. The degree of the factor runs counter to the number of differentiations: the fewer the differentiations, the higher the degree. This progression reaches its maximum for the term with the lowest nonzero derivative, namely $F_n^{(1)}(z)$, where the degree is ${(n-1)}$. However, for the term with no derivative, $F_n(z)$, the degree is not maximal but drops by one, giving ${(n-2)}$. The corresponding coefficients $\mathcal{Z}$ are set to zero, although this is not immediately obvious from the definition.

Having constructed the general differential equations of the form (\ref{15}) or (\ref{16}), we now verify them using standard CFT methods. Next, we discuss the algorithm for obtaining the BPZ equations without a detailed introduction to CFT; further theoretical aspects can be found in \cite{belavin1984infinite, zamolodchikov1989conformal}. The correlation function with one degenerate field $\Phi_{n,1}$ satisfies a differential equation, which follows from the vanishing of correlation function of the following form:
\begin{align} \label{20}
    %\langle \Phi_{n,1}(z) \Phi_1(z_1) ... \Phi_n(z_n) \rangle = 
    \langle \mathcal{D}_{n,1} \Phi_{n,1}(z) \Phi_1(z_1) \dots \Phi_N(z_N) \rangle=0,
\end{align}
where $\mathcal{D}_{n,1}$ is the "singular vector creation operator". $\mathcal{D}_{n,1}$ is defined as a specific combination of the Virasoro operators $L_{-k}$, and the operators for nontrivial cases with $n\leq 9$ are given in \hyperref[sec: Apendix B]{Appendix A}. Correlation functions involving the field $L_{-k} \Phi(z)$ are calculated according to the following rule:
\begin{align} \label{21}
    \langle L_{-k} \Phi(z) \Phi_1(z_1) ... \Phi_N(z_N) \rangle =& \sum_{j=1}^N \bigg( \frac{\Delta_j (k-1)}{(z_j-z)^k} - \frac{\partial_j}{(z_j-z)^{k-1}} \bigg) \langle \Phi(z) \Phi_1(z_1) ... \Phi_N(z_N) \rangle, 
\end{align}
where $\Delta_j = \alpha_j(Q - \alpha_j)$ is the conformal dimension of the field $\Phi_j$. From (\ref{20}) and (\ref{21}), one readily observes that, due to the form of $\mathcal{D}_{n,1}$, a specific combination of differential operators is applied to the correlation function. In this paper, we restrict our attention to the four-point correlation function $f_n(z) = \langle \Phi_{n,1}(z) \Phi_1(z_1) \Phi_2(z_2) \Phi_3(z_3) \rangle$. By conformal invariance, the three points can be fixed as $(z, z_1, z_2, z_3) = (z, 0, 1, \infty)$. After this fixing and the substitution $f_n(z) = z^{(n-1) b \alpha_1} (1 - z)^{(n-1) b \alpha_2} F_n(z)$, the differential operator reduces to a linear ordinary differential equation of order $n$ in the variable $z$. The result coincides exactly with the general form (\ref{16}) with the parameters (\ref{n2}). The comparison was performed for equations up to the ninth order. For the case of the degenerate fields $\Phi_{1,n}$, the same result follows by the duality transformation $b \to b^{-1}$.

\section{Connection with Hypergeometric operators}
The generalized hypergeometric series $_{p}F_q$ is defined as
\begin{align}
    _{p}F_q(a_1,\dots,a_p;b_1,\dots,b_q;z) = \sum_{k=0}^{\infty} \frac{(a_1)^{\overline{k}} \dots (a_p)^{\overline{k}}}{(b_1)^{\overline{k}} \dots (b_q)^{\overline{k}}} \frac{z^k}{k!}, \hspace{25pt} |z|<1 , \hspace{10pt} p \leq q+1.
\end{align}
It satisfies the generalized hypergeometric differential equation $_pE_q$
\begin{align}
    \Big(\theta(\theta + b_1 - 1) \cdot  ... \cdot (\theta + b_q - 1) - z (\theta + a_1)  \cdot  ... \cdot (\theta + a_p) \Big) y(z)=0,
\end{align}
where $\theta = z \partial_z$. We consider the hypergeometric differential operators, including parameter values for which the corresponding hypergeometric series may not be well-defined. 

In this section, we focus on the case of ${}_{p+1}E_{p}$, which has regular singular points $z=0,1,\infty$. After converting this equation from the operator $(\theta,\partial)$-notation to the normal $(z,\partial)$-form, it can be written schematically as follows:
\begin{align} \label{19}
    (z-1) z^{p} y^{(p+1)}(z) + \sum_{k=1}^p (f_{b,k} + f_{a,k} z) z^{k-1}y^{(k)}(z) + \prod_{k=1}^{p+1} a_k y(z) =0,
\end{align}
where $f_{b,k}$ and $f_{a,k}$ depend on the parameters $(a_1,\dots,a_{p+1};b_1,\dots,b_p)$. It is easy to notice that a similar equation can be obtained from (\ref{16}) by setting some of the parameters $\mathcal{Z}_k(l,m,0)$ to zero. Thus, the candidate for hypergeometric operator takes the form:
\begin{align} \label{n20}
   \sum_{k=-1}^{p} ((z-1)\mathcal{Z}_{p+1-k}(p - k,0,0) + z\mathcal{Z}_{p+1-k}(p - k-1,1,0)) (-1)^{k} z^{k}F_{p+1}^{(k+1)}(z) = 0,
\end{align}
where $\mathcal{Z}_k(l,m,i)=0$ for $l<0$ or $m<0$, and vanishing factors 
\begin{align} \label{n21}
    \{ \mathcal{Z}_k(k-3,2,0),\dots,\mathcal{Z}_k(0,k-1,0) \} = 0, \hspace{20pt} 3 \leq k \leq p+2
\end{align}
impose constraints on the parameters $\{ A,B,C,g \}$.

Let us consider the correspondence between (\ref{19}) and (\ref{n20}). As mentioned in \cite{mimachi2024irreducibility}, the condition $2B+2C+g+1=0$ allows us to represent the Dotsenko--Fateev equation (\ref{1}) as a hypergeometric operator. Moreover, the system (\ref{n21}) has other restrictions: $\{ B + C = 0, 2 + 2 A + g = 0 \}$ and $\{ B = 0, C = 0 \}$. These conditions are more rigid because they restrict more parameters. 

The next case turns out to be more complicated. There are conditions that clearly fix 2, 3 or 4 of the parameters $\{ A, B, C, g \}$. At the same time, the case of fixing two parameters is accompanied by a condition on the remaining two. All results for third- and fourth-order differential equations are presented in \hyperref[sec: Apendix A]{Appendix B}.

A more interesting and natural situation occurs when we consider higher-order differential equations. For the orders examined up to $n=9$, the constraints (29) yield two families of restrictions that do not completely fix the parameters $\{A,B,C,g\}$: $\{B=0,C=0,g=0\}$ and $\{A=-1,g=0,B+C=0\}$. This suggests that the same pattern may persist at higher orders. The corresponding hypergeometric parameters can be written in the following general form:

%\newpage
\begin{table}[h]
\centering
\begin{tabular}{|C{5cm}|C{5cm}|}
\hline
Restrictions & Parameters \\ % & — разделители между ячейками \\ — конец строки
\hline
$B = 0, C = 0, g = 0$ & 
$a_k = - k (1 + A)$,  $a_n = 0$ \newline $b_k = -(k-1) - k A$ \\
\hline
$A = -1, B = -C, g = 0$ & 
$a_k = - k C$, $a_n = 0$ \newline $b_k = 1 - k C$ \\
\hline
\end{tabular}
%\caption{Caption}
%\label{tab:placeholder}
\end{table}
\noindent where $1\leq k\leq n-1$. It is important to mention that such restrictions also appear in the fourth-order case.

\section{Conclusion}
In this paper, we have constructed differential equations of arbitrary order for Dotsenko–Fateev integrals. We have presented an algorithm for reconstructing the differential equation from an integral solution. The general structure of such equations has been found, and their coefficients are expressed via recurrence relations, making the construction fully algorithmic.

The resulting equations have been verified on special cases arising in CFT. In particular, we have verified up to the ninth order that the BPZ equations are special cases of our general equations. Thus, our approach provides an independent check of the correspondence between Dotsenko–Fateev integrals and BPZ equations in these cases. It is worth noting that the standard derivation of the BPZ equations is technically quite involved. In this respect, our method is more efficient, since it allows one to obtain differential equations in a uniform way and with a lower risk of calculation errors.

Furthermore, the general form of equations has been rewritten in a canonical form, which made it possible to establish a direct connection with hypergeometric operators. This observation may be useful for studying deeper connections between these constructions.

Finally, we believe that the developed approach can be extended to supersymmetric generalizations of CFT. At the same time, it would be natural to generalize the construction to degenerate fields $\Phi_{n,m}$ with arbitrary indices, beyond the $\Phi_{n,1}$ and $\Phi_{1,n}$ cases considered here. The study of analogous differential equations in these contexts represents a natural direction for our future research.

\begin{comment}
In this paper, we have considered a special class of higher-order differential equations. We have studied the algorithm for reconstructing the differential equation from its solution and generalized it. The general form for higher-order equations corresponding to the Dotsenko–Fateev integrals was obtained. The coefficients of these equations satisfy the recurrence relations. Furthermore, the general form of differential equations have been verified against the BPZ equations in CFT, and the verification has been carried out up to and including the ninth order. 

Furthermore, the differential equation was reduced to canonical form, which allowed us to compare it with the well-known class of hypergeometric equations. Cases in which the equations reduce to hypergeometric ones were identified. This correspondence may be useful in the future study of the obtained operators, since some of their properties may be inherited from the hypergeometric ones.
\end{comment}

\section*{Acknowledgments}
I would like to thank Alexey Litvinov for valuable discussions.

\section*{Appendix A} \label{sec: Apendix B}
In this appendix, we list some of the operators $\mathcal{D}_{n,1}$ that are used in our framework for the construction of differential equations.
\begin{align*}
    &\mathcal{D}_{2,1} = L_{-1}^2 + b^2 L_{-2}, \hspace{30pt} \\
    &\mathcal{D}_{3,1} = L_{-1}^3 + 4 b^2 L_{-2} L_{-1} + 2 b^2 (1 + 2 b^2) L_{-3}, \\
    &\mathcal{D}_{4,1} = L_{-1}^4 + 10 b^2 L_{-2} L_{-1}^2 + 9 b^4 L_{-2}^2 + 2 b^2 (5 + 12 b^2) L_{-3} L_{-1} + 6 b^2 (1 + 4 b^2 + 6 b^4) L_{-4}, \\
    &\mathcal{D}_{5,1} = L_{-1}^5 + 20 b^2 L_{-2} L_{-1}^3 + 64 b^4 L_{-2}^2 L_{-1} + 6 b^2 (5 + 14 b^2) L_{-3} L_{-1}^2 + 64 b^4 (1 + 3 b^2) L_{-3} L_{-2} + \\
    &\hspace{35pt} + 12 b^2 (3 + 14 b^2 + 24 b^4) L_{-4} L_{-1} + 8 b^2 (1 + 3 b^2) (3 + 14 b^2 + 24 b^4) L_{-5}, \\
    &\mathcal{D}_{6,1} = L_{-1}^6 + 35 b^2 L_{-2} L_{-1}^4 + 259 b^4 L_{-2}^2 L_{-1}^2 + 225 b^6 L_{-2}^3 + 14 b^2 (5 + 16 b^2) L_{-3} L_{-1}^3 + \\
    &\hspace{35pt} + 2 b^4 (259 + 880 b^2) L_{-3} L_{-2} L_{-1} + 10 b^4 (13 + 88 b^2 + 160 b^4) L_{-3}^2 + 6 b^2 (21 + 112 b^2 + 216 b^4) L_{-4} L_{-1}^2 + \\
    &\hspace{35pt} + 10 b^4 (31 + 176 b^2 + 360 b^4) L_{-4} L_{-2} + 4 b^2 (42 + 369 b^2 + 1192 b^4 + 1440 b^6) L_{-5} L_{-1} + \\
    &\hspace{35pt} + 40 b^2 (3 + 35 b^2 + 160 b^4 + 378 b^6 + 360 b^8) L_{-6}, \\
    &\mathcal{D}_{7,1} = L_{-1}^7 + 56 b^2 L_{-2} L_{-1}^5 + 784 b^4 L_{-2}^2 L_{-1}^3 + 2304 b^6 L_{-2}^3 L_{-1} + 28 b^2 (5 + 18 b^2) L_{-3} L_{-1}^4 + \\
    &\hspace{35pt} + 48 b^4 (49 + 186 b^2) L_{-3} L_{-2} L_{-1}^2 + 3456 b^6 (1 + 4 b^2) L_{-3} L_{-2}^2 + 4 b^4 (295 + 2232 b^2 + 4500 b^4) L_{-3}^2 L_{-1} + \\
    &\hspace{35pt} + 48 b^2 (7 + 42 b^2 + 90 b^4) L_{-4} L_{-1}^3 + 64 b^4 (44 + 279 b^2 + 630 b^4) L_{-4} L_{-2} L_{-1} + \\
    &\hspace{35pt} + 24 b^4 (1 + 4 b^2) (59 + 360 b^2 + 900 b^4) L_{-4} L_{-3} + 32 b^2 (21 + 208 b^2 + 747 b^4 + 990 b^6) L_{-5} L_{-1}^2 + \\
    &\hspace{35pt} + 144 b^4 (1 + 4 b^2) (13 + 84 b^2 + 180 b^4) L_{-5} L_{-2} + 160 b^2 (1 + 6 b^2) (6 + 43 b^2 + 144 b^4 + 180 b^6) L_{-6} L_{-1} + \\
    &\hspace{35pt} + 120 b^2 (1 + 4 b^2) (1 + 6 b^2) (6 + 43 b^2 + 144 b^4 + 180 b^6) L_{-7} \\
    &\mathcal{D}_{8,1} = L_{-1}^8 + 84 b^2 L_{-2} L_{-1}^6 + 1974 b^4 L_{-2}^2 L_{-1}^4 + 12916 b^6 L_{-2}^3 L_{-1}^2 + 11025 b^8 L_{-2}^4 + 252 b^2 (1 + 4 b^2) L_{-3} L_{-1}^5 + \\
    &\hspace{35pt} + 24 b^4 (329 + 1380 b^2) L_{-3} L_{-2} L_{-1}^3 + 12 b^6 (3229 + 14196 b^2) L_{-3} L_{-2}^2 L_{-1} + \\
    &\hspace{35pt} + 36 b^4 (165 + 1380 b^2 + 3056 b^4) L_{-3}^2 L_{-1}^2 + 56 b^6 (347 + 3042 b^2 + 7056 b^4) L_{-3}^2 L_{-2} + \\
    &\hspace{35pt} + 36 b^2 (21 + 140 b^2 + 330 b^4) L_{-4} L_{-1}^4 + 72 b^4 (197 + 1380 b^2 + 3410 b^4) L_{-4} L_{-2} L_{-1}^2 + \\
    &\hspace{35pt} + 28 b^6 (827 + 6084 b^2 + 15750 b^4) L_{-4} L_{-2}^2 + 48 b^4 (297 + 3316 b^2 + 14283 b^4 + 23940 b^6) L_{-4} L_{-3} L_{-1} + \\
    &\hspace{35pt} + 84 b^4 (51 + 712 b^2 + 4380 b^4 + 13680 b^6 + 18900 b^8) L_{-4}^2 + \\
    &\hspace{35pt} + 48 b^2 (42 + 463 b^2 + 1830 b^4 + 2640 b^6) L_{-5} L_{-1}^3 + \\
    &\hspace{35pt} + 16 b^4 (1179 + 13649 b^2 + 56682 b^4 + 85680 b^6) L_{-5} L_{-2} L_{-1} + \\
    &\hspace{35pt} + 14 b^4 (678 + 10663 b^2 + 66936 b^4 + 200304 b^6 + 241920 b^8) L_{-5} L_{-3} + \\
    &\hspace{35pt} + 240 b^2 (18 + 264 b^2 + 1480 b^4 + 4191 b^6 + 4680 b^8) L_{-6} L_{-1}^2 + \\
    &\hspace{35pt} + 112 b^4 (120 + 1847 b^2 + 10884 b^4 + 32355 b^6 + 37800 b^8) L_{-6} L_{-2} + \\
    &\hspace{35pt} + 6 b^2 (1080 + 20670 b^2 + 159031 b^4 + 641472 b^6 + 1374480 b^8 + 1209600 b^{10}) L_{-7} L_{-1} + \\
    &\hspace{35pt} + 84 b^2 (60 + 1390 b^2 + 13607 b^4 + 72568 b^6 + 229128 b^8 + 400320 b^{10} + 302400 b^{12}) L_{-8}, 
\end{align*}
\begin{align*}
    &\mathcal{D}_{9,1} = L_{-1}^9 + 120 b^2 L_{-2} L_{-1}^7 + 4368 b^4 L_{-2}^2 L_{-1}^5 + 52480 b^6 L_{-2}^3 L_{-1}^3 + 147456 b^8 L_{-2}^4 L_{-1} + \\
    &\hspace{35pt} + 84 b^2 (5 + 22 b^2) L_{-3} L_{-1}^6 + 240 b^4 (91 + 418 b^2) L_{-3} L_{-2} L_{-1}^4 + 384 b^6 (615 + 2948 b^2) L_{-3} L_{-2}^2 L_{-1}^2 + \\
    &\hspace{35pt} + 294912 b^8 (1 + 5 b^2) L_{-3} L_{-2}^3 + 12 b^4 (1825 + 16720 b^2 + 40348 b^4) L_{-3}^2 L_{-1}^3 + \\
    &\hspace{35pt} + 256 b^6 (925 + 8844 b^2 + 22260 b^4) L_{-3}^2 L_{-2} L_{-1} + 256 b^6 (1 + 5 b^2) (155 + 1442 b^2 + 3920 b^4) L_{-3}^3 + \\
    &\hspace{35pt} + 72 b^2 (21 + 154 b^2 + 396 b^4) L_{-4} L_{-1}^5 + 480 b^4 (109 + 836 b^2 + 2244 b^4) L_{-4} L_{-2} L_{-1}^3 + \\
    &\hspace{35pt} + 256 b^6 (1103 + 8844 b^2 + 24768 b^4) L_{-4} L_{-2}^2 L_{-1} + \\
    &\hspace{35pt} + 24 b^4 (3285 + 40172 b^2 + 188364 b^4 + 341712 b^6) L_{-4} L_{-3} L_{-1}^2 + \\
    &\hspace{35pt} + 1024 b^6 (1 + 5 b^2) (277 + 2154 b^2 + 6552 b^4) L_{-4} L_{-3} L_{-2} + \\
    &\hspace{35pt} + 48 b^4 (987 + 15092 b^2 + 101060 b^4 + 341712 b^6 + 508032 b^8) L_{-4}^2 L_{-1} + \\
    &\hspace{35pt} + 240 b^2 (21 + 255 b^2 + 1100 b^4 + 1716 b^6) L_{-5} L_{-1}^4 + \\
    &\hspace{35pt} + 16 b^4 (6525 + 82820 b^2 + 373692 b^4 + 608400 b^6) L_{-5} L_{-2} L_{-1}^2 + \\
    &\hspace{35pt} + 3072 b^6 (1 + 5 b^2) (61 + 504 b^2 + 1296 b^4) L_{-5} L_{-2}^2 + \\
    &\hspace{35pt} + 8 b^4 (13125 + 226308 b^2 + 1546828 b^4 + 5004528 b^6 + 6491520 b^8) L_{-5} L_{-3} L_{-1} + \\
    &\hspace{35pt} + 64 b^4 (1 + 5 b^2) (987 + 15092 b^2 + 101060 b^4 + 341712 b^6 + 508032 b^8) L_{-5} L_{-4} + \\
    &\hspace{35pt} + 480 b^2 (30 + 485 b^2 + 2970 b^4 + 9108 b^6 + 10920 b^8) L_{-6} L_{-1}^3 + \\
    &\hspace{35pt} + 64 b^4 (2325 + 39264 b^2 + 251588 b^4 + 806544 b^6 + 1008000 b^8) L_{-6} L_{-2} L_{-1} + \\
    &\hspace{35pt} + 128 b^4 (1 + 5 b^2) (585 + 9616 b^2 + 62292 b^4 + 208656 b^6 + 282240 b^8) L_{-6} L_{-3} + \\
    &\hspace{35pt} + 48 b^2 (675 + 14250 b^2 + 119943 b^4 + 524636 b^6 + 1208628 b^8 + 1134000 b^{10}) L_{-7} L_{-1}^2 + \\
    &\hspace{35pt} + 256 b^4 (1 + 5 b^2) (435 + 7412 b^2 + 47324 b^4 + 149472 b^6 + 181440 b^8) L_{-7} L_{-2} + \\
    &\hspace{35pt} + 16 b^2 (3150 + 80535 b^2 + 863450 b^4 + 5000652 b^6 + 17001880 b^8 + 31721760 b^{10} + 25401600 b^{12}) L_{-8} L_{-1} + \\
    &\hspace{35pt} + 64 b^2 (1 + 5 b^2) (630 + 16107 b^2 + 171698 b^4 + 985372 b^6 + 3375288 b^8 + 6344352 b^{10} + 5080320 b^{12}) L_{-9} 
\end{align*}

\section*{Appendix B} \label{sec: Apendix A}
In this appendix, we present all restrictions that allow the third- and fourth-order equations of the Dotsenko--Fateev type to be represented as hypergeometric operators, together with the corresponding set of hypergeometric parameters.

\subsection*{Third-order equation}
\begin{table}[h]
\centering
\begin{tabular}{|C{5cm}|C{9cm}|} % здесь в скобках указываются количество столбцов, их параметры и вертикальные линии
\hline % горизонтальная линия (верхняя)
Restrictions & Parameters \\ % & — разделители между ячейками \\ — конец строки
\hline % горизонтальная линия (после первой строки)
$2 B + 2 C + g + 1 = 0$ & $a_1 = -1 - 2 A$, $a_2 = -A + B$, $a_3 = -2 C$, \newline $ b_1 = -2 A + 2 B$, $b_2 = -A - C,$ \\
\hline % горизонтальная линия (после второй строки)
$B + C = 0$, $2 + 2 A + g = 0$ & $a_1 = 2 B$, $a_2 = 1 + A + B$, $a_3 = 0$, \newline $b_1 = -A + B$, $b_2 = 1 + 2 B$ \\
\hline
$B = 0$, $C = 0$ & $a_1 = -1 - A - g$, $a_2 = -2 - 2 A - g$, $a_3 = 0$, \newline $b_1 = -A$, $b_2 = -1 - 2 A - g$ \\
\hline % горизонтальная линия (после последней строки)
\end{tabular}
\end{table}

\subsection*{Fourth-order equation}
\begin{tabular}{|C{6.5cm}|C{10cm}|} % здесь в скобках указываются количество столбцов, их параметры и вертикальные линии
\hline % горизонтальная линия (верхняя)
Restrictions & Parameters \\ % & — разделители между ячейками \\ — конец строки
\hline % горизонтальная линия (после первой строки)1
$A = -1$, $B = -C$, $g = 0$ & $a_1 = 0$, $a_2 = -C$, $a_3 = - 2C$, $a_4 = - 3C$, \newline $b_1 = 1 - C$, $b_2 = 1 - 2C$, $b_3 = 1 - 3C$ \\
\hline % горизонтальная линия (после второй строки)2
$A = - 1/3$, $B = 1 - C$, $g = -5/3$ & $a_1 = 0$, $a_2 =5/3 -C$, $a_3 =5/3 - 2C$, $a_4 = - 3C$, \newline $b_1 = 1/3 - C$, $b_2 = 4/3 - 2C$, $b_3 = 4 - 3C$ \\
\hline % горизонтальная линия (после последней строки)3
$A = 2/3$, $B = 1 - C$, $g = -5/3$ & $a_1 = -3$, $a_2 =-1/3 -C$, $a_3 = 2/3 - 2C$, $a_4 = - 3C$, \newline $b_1 = -2/3 - C$, $b_2 = -2/3 - 2C$, $b_3 = 1 - 3C$ \\
\hline % горизонтальная линия (после последней строки)4
$A = -2/3$, $B = -C$, $g = -1/3$ & $a_1 = 0$, $a_2 =1/3 -C$, $a_3 = 1/3 - 2C$, $a_4 = - 3C$, \newline $b_1 = 2/3 - C$, $b_2 = 2/3 - 2C$, $b_3 = 1 - 3C$ \\
\hline % горизонтальная линия (после последней строки)5
$A = -1$, $B = -2/3 - C$, $g = 1/3$ & $a_1 = 1$, $a_2 =1/3 -C$ , $a_3 = - 2C$ , $a_4 = - 3C$, \newline $b_1 = 1 - C$ , $b_2 = 2/3 - 2C$ , $b_3 = - 3C$ \\
\hline % горизонтальная линия (после последней строки)6
$B = 0$, $C = 0$, $g = 0$ & $a_1 = 0 $, $a_2 = -1 -A$, $a_3 = -2 -2A$, $a_4 = -3 -3A$, \newline $b_1 = -A$, $b_2 = -1 -2A$, $b_3 = -2-3A$ \\
\hline % горизонтальная линия (после последней строки)7
$B = -1/3$, $C = -1/3$, $g = 1/3$ & $a_1 = 1$, $a_2 = -1/3 - A$, $a_3 = -4/3 - 2A$, $a_4 = -2 - 3A$, \newline $b_1 = 1/3 - A$, $b_2 = -2/3 - 2A$, $b_3 = -2 - 3A$ \\
\hline % горизонтальная линия (после последней строки)8
$B = 1$, $C = 0$, $g = -5/3$ & $a_1 = 0$, $a_2 = 4/3 - A$, $a_3 = 1 - 2A$, $a_4 = -1 - 3A$, \newline $b_1 = - A$, $b_2 = 2/3 - 2A$, $b_3 = 3 - 3A$ \\
\hline % горизонтальная линия (после последней строки)9
$B = 0$, $C = 0$, $g = -1/3$ & $a_1 = 0$, $a_2 = -1/3 - A$, $a_3 = -1 - 2A$, $a_4 = -2 - 3A$, \newline $b_1 = - A$, $b_2 = -2/3 - 2A$, $b_3 = -1 - 3A$ \\
\hline % горизонтальная линия (после последней строки)10
$B = 0$, $C = 1$, $g = -5/3$ & $a_1 = -3$, $a_2 = -2/3 - A$, $a_3 = - 2A$, $a_4 = -1 - 3A$, \newline $b_1 = -1 - A$, $b_2 = -4/3 - 2A$, $b_3 = - 3A$ \\
\hline % горизонтальная линия (после последней строки)11
$A = -2/3$, $B = -1/3$, $C = -1/3$, $g = 1/3$ & $a_1 = 0$, $a_2 = 0$, $a_3 = 1/3$, $a_4 = 1$, \newline $b_1 = 0$, $b_2 = 2/3$, $b_3 = 1$ \\
\hline % горизонтальная линия (после последней строки)12
$A = 3/2$, $B = 1$, $C = 0$, $g = -3$ & $a_1 = 0$, $a_2 = -3/2$, $a_3 = 2$, $a_4 = 5/2$, \newline $b_1 = -3/2$, $b_2 = -1$, $b_3 = 5/2$ \\
\hline % горизонтальная линия (после последней строки)13
$A = 0$, $B = 2/3$, $C = 0$, $g = -4/3$ & $a_1 = 0$, $a_2 = -1$, $a_3 = 2/3$, $a_4 = 1$, \newline $b_1 = 0$, $b_2 = 1/3$, $b_3 = 2$ \\
\hline % горизонтальная линия (после последней строки)14
$A = -1$, $B = -2/3$, $C = 0$, $g = 1/3$ & $a_1 = 0$, $a_2 = 0$, $a_3 = 1/3$, $a_4 = 1$, \newline $b_1 = 0$, $b_2 = 2/3$, $b_3 = 1$ \\
\hline % горизонтальная линия (после последней строки)15
$A = -1/3$, $B = 2/3$, $C = 1/3$, $g = -5/3$ & $a_1 = 0$, $a_2 = -1$, $a_3 = 1$, $a_4 = 4/3$, \newline $b_1 = 0$, $b_2 = 2/3$, $b_3 = 3$ \\
\hline % горизонтальная линия (после последней строки)16
$A = 2/3$, $B = 2/3$, $C = 1/3$, $g = -5/3$ & $a_1 = 0$, $a_2 = -3$, $a_3 = -1$, $a_4 = -2/3$, \newline $b_1 = 0$, $b_2 = -4/3$, $b_3 = -1$ \\
\hline % горизонтальная линия (после последней строки)17
$A = -1/3$, $B = 1/3$, $C = 1/3$, $g = -4/3$ & $a_1 = 0$, $a_2 = -1$, $a_3 = 2/3$, $a_4 = 1$, \newline $b_1 = 0$, $b_2 = 1/3$, $b_3 = 2$ \\
\hline % горизонтальная линия (после последней строки)18
$A = 1/3$, $B = 1/3$, $C = 1/3$, $g = -4/3$ & $a_1 = 0$, $a_2 = -2$, $a_3 = -1$, $a_4 = -1/3$, \newline $b_1 = 0$, $b_2 = -1$, $b_3 = -2/3$ \\
\hline % горизонтальная линия (после последней строки)19
$A = 1$, $B = 1/2$, $C = 1/2$, $g = -3$ & $a_1 = 0$, $a_2 = -3/2$, $a_3 = 2$, $a_4 = 5/2$, \newline $b_1 = -3/2$, $b_2 = -1$, $b_3 = 5/2$ \\
\hline % горизонтальная линия (после последней строки)20
$A = 2$, $B = 1/2$, $C = 1/2$, $g = -3$ & $a_1 = -3$, $a_2 = -3/2$, $a_3 = 1/2$, $a_4 = 1$, \newline $b_1 = -3$, $b_2 = -5/2$, $b_3 = -1/2$ \\
\hline
\end{tabular}

\begin{tabular}{|C{6.5cm}|C{10cm}|} % здесь в скобках указываются количество столбцов, их параметры и вертикальные линии
\hline % горизонтальная линия (верхняя)
Restrictions & Parameters \\ % & — разделители между ячейками \\ — конец строки
\hline % горизонтальная линия (после последней строки)21
$A = -1$, $B = -7/6$, $C = 1/2$, $g = 1/3$ & $a_1 = -3/2$, $a_2 = -1$, $a_3 = -1/6$, $a_4 = 1$, \newline $b_1 = -3/2$, $b_2 = -1/3$, $b_3 = 1/2$ \\
\hline % горизонтальная линия (после последней строки)22
$A = 2/3$, $B = 1/3$, $C = 2/3$, $g = -5/3$ & $a_1 = -3$, $a_2 = -2$, $a_3 = -1$, $a_4 = -2/3$, \newline $b_1 = -2$, $b_2 = -4/3$, $b_3 = -1$ \\
\hline % горизонтальная линия (после последней строки)23
$A = 0$, $B = 0$, $C = 2/3$, $g = -4/3$ & $a_1 = 0$, $a_2 = -2$, $a_3 = -1$, $a_4 = -1/3$, \newline $b_1 = 0$, $b_2 = -1$, $b_3 = -2/3$ \\
\hline % горизонтальная линия (после последней строки)24
$A = 3/2$, $B = 0$, $C = 1$, $g = -3$ & $a_1 = -3$, $a_2 = -3/2$, $a_3 = 1/2$, $a_4 = 1$, \newline $b_1 = -3$, $b_2 = -5/2$, $b_3 = -1/2$ \\
\hline % горизонтальная линия (после последней строки)25
$A = -1$, $B = -5/3$, $C = 1$, $g = 1/3$ & $a_1 = -3$, $a_2 = -2$, $a_3 = -2/3$, $a_4 = 1$, \newline $b_1 = 0$, $b_2 = -3$, $b_3 = -4/3$ \\
\hline
\end{tabular}

\bibliographystyle{MyStyle}
\bibliography{bibliography}

\end{document}